\documentclass[twocolumn,aps,prb,showpacs]{revtex4-1}

\usepackage{graphicx}
\usepackage{epstopdf}
\usepackage{amsmath}
\usepackage{amssymb}
\usepackage{upgreek}
\usepackage[tight,TABTOPCAP]{subfigure}
\usepackage{float}
\usepackage{wrapfig}
\usepackage{natbib}
\usepackage{color}

\makeatletter

\begin{document}

\title{Impact of Electron-Phonon Coupling on Near-Field Optical Spectra}

\author{J. P. Carbotte$^{1,2}$}
\author{J. P. F. LeBlanc$^{3}$}
\email{jpfleblanc@gmail.com}
\author{Phillip E. C. Ashby$^{1}$}

\affiliation{$^1$Department of Physics and Astronomy, McMaster
University, Hamilton, Ontario L8S~4M1 Canada}
\affiliation{$^2$The Canadian Institute for Advanced Research, Toronto, ON M5G~1Z8 Canada}
\affiliation{$^3$Max-Planck-Institute for the Physics of Complex Systems, 01187 Dresden, Germany}


\date{\today}
\begin{abstract}
The finite momentum transfer ($\boldsymbol{q}$) longitudinal optical response $\sigma^L(\boldsymbol{q},\omega)$ of graphene has a peak at an energy $\omega=\hbar v_F q$.  This corresponds directly to a quasiparticle peak in the spectral density at momentum relative to the Fermi momentum $k_F -q$.  Inclusion of coupling to a phonon mode at $\omega_E$ results, for $\omega<|\omega_E|$, in an electron-phonon renormalization of the bare bands by a mass enhancement factor $(1+\lambda)$ and this is followed by a phonon kink for $\omega$ around $\omega_E$ where additional broadening begins.  Here we study the corresponding changes in the optical quasiparticle peaks which we find continues to  track directly the renormalized quasiparticle energies until $q$ is large enough that the optical transitions begin to sample the phonon kink region of the dispersion curves where linearity in momentum  and the correspondence to a single quasiparticle energy are lost.  Nevertheless there remains in $\sigma^L(\boldsymbol{q},\omega)$ features analogous to the phonon kinks of the dispersion curves which are observable through variation of $q$ and $\omega$.
\end{abstract}

\maketitle

Important information on the charge dynamics of the Dirac fermions in graphene is obtained in optical absorption experiments.  Results are reviewed by Orlita and Potemski.\cite{orlita:2010}  For bare bands the real part of the zero momentum limit optical conductivity, $\sigma(\boldsymbol{q}=0,\omega)$, has a Drude peak around $\omega=0$ which has its origin in the intraband optical transitions.
There is also an additional piece due to the interband transitions with onset at twice the chemical potential which provides a constant universal background of value $\sigma_0=\pi e^2/2h$.\cite{gusynin:2007,ando:2002,gusynin:2009}  While in the clean limit the Pauli blocked region between the Drude and the universal background would have essentially no conductivity, experimental work\cite{li:2008} has found instead a value of order $\sigma_0/3$.  This observation can be partially explained \cite{peres:2008,stauber:2008b,stauber:2008,carbotte:2010}
due to impurities, electron-phonon interactions and/or coulomb correlations, although its precise origin remains controversial.  
In conventional metals the electron-phonon interaction is known to renormalize many of their properties.\cite{prange:1964}
It leads  to incoherent phonon-assisted Holstein side bands\cite{mori:2008} in addition to the main coherent Drude response with optical spectral weight now being distributed between these two parts.  The effect on the Drude weight can be expressed in terms of the electron phonon mass renormalization parameter, $\lambda$. The Drude weight is reduced with a multiplicative factor of $1/(1+\lambda)$ while the remaining $\lambda/(1+\lambda)$ is transferred to the sideband due to the new absorption processes in which a phonon is created by a photon along with a hole-particle pair.  In graphene, electron-phonon effects have been seen\cite{li:2009} in the electronic density of states (DOS), $N(\omega)$, measured in scanning tunnelling spectroscopy (STS).  These observations are expected in systems where the DOS is energy dependent.\cite{mitrovic:1983,mitrovic:1983:2,cappelluti:2003,nicol:2009}
Phonon `kinks' have also been seen in angle-resolved photoemission spectroscopy (ARPES) \cite{bostwick:2007,zhou:2008} as predicted theoretically.\cite{calandra:2007, park:2009, park:2009:nl, tse:2007}
Recently, near-field optical techniques have been used to obtain information on the finite momentum conductivity \cite{fei:2011,fei:2012,chen:2012,carbotte:2012}
rather than the standard long wavelength, $q\to0$, limit.  This has allowed for  nano-imaging\cite{fei:2011,fei:2012,chen:2012} of the graphene plasmons and could in principle be employed to get information on plasmarons\cite{carbotte:2012} a scattering resonance of an electron and a plasmon.\cite{lundqvist:1967}  In this paper we show how $\sigma(\boldsymbol{q},\omega)$ can provide information on electron-phonon renormalization effects complementary to ARPES and STM.

\begin{figure}
  \begin{center}
  \includegraphics[width=0.95\linewidth]{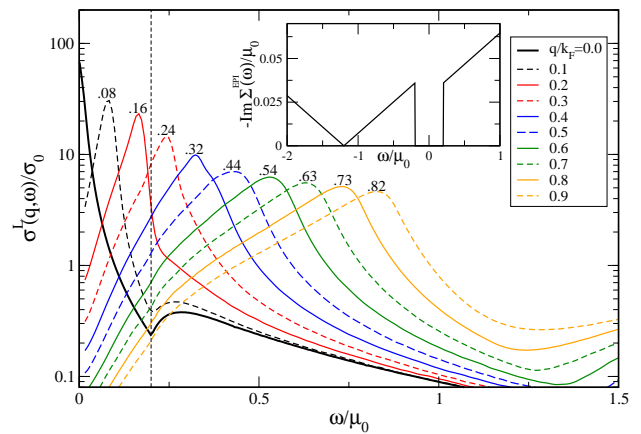}
  \end{center}
  \caption{\label{fig:1}(Color online) The real part of the finite momentum longitudinal optical conductivity, $\sigma^L(q,\omega)/\sigma_0$ as a function of $\omega/\mu_0$ for a selection of values of $q/k_F$ from $0.0\to 0.9$ (as defined in legend). These include an electron-phonon interaction with $\lambda\approx0.18$ for a single optical phonon mode at $\omega_E=200$~meV (shown by vertical dashed line), where here $\mu_0=1$~eV, ${\rm A}=0.08$~eV, $W_c=7.0$~eV and include a residual scattering $\eta=0.005\mu_0$. Inset: $-{\rm Im}\Sigma^{EPI}(\omega)/\mu_0$ as in Eq.~(\ref{eqn:imsigma}) for parameters in the main frame.  }
\end{figure}  

The Kubo formula for the real part of the finite momentum optical conductivity is approximated by the simplest bubble diagram which is not exact since it neglects vertex corrections. This includes interactions through a self energy and is given by
\begin{align}
&\frac{\sigma(\boldsymbol{q},\omega)}{\sigma_0} =\frac{8}{\omega}\int\limits_{-\omega}^0 d \omega^\prime \nonumber \\ &\times\int \frac{d^2\boldsymbol{k}}{2\pi} \sum_{s,s^\prime=\pm} F_{ss^\prime}(\phi) A^s(\boldsymbol{k},\omega^\prime)A^{s^\prime}(\boldsymbol{k}+\boldsymbol{q},\omega^\prime +\omega),
 \label{eqn:cond}
\end{align}
where $A^s(k,\omega)$ is the spectral function of the Dirac fermions.  Cappelluti and Benfatto\cite{cappelluti:2009} have studied the effect of vertex corrections on the conductivity of graphene and found that, as is the case in conventional metals, their main effect can be incorporated into our Eqn.~(\ref{eqn:cond}) by changing the scattering rate from its quasiparticle value to an appropriate optical rate which contains an extra factor of $(1-\cos\beta)$ where $\beta$ is the scattering angle.  This well known factor gives more weight to backward scattering and de-emphasises forward scattering.  The overlap factor, $F_{ss^\prime}(\phi)$, has the form $F_{ss^\prime}(\phi)=\frac{1}{2}[1+ss^\prime \cos\phi]$ where $\phi$ is the sum of the angles of initial momentum, $\boldsymbol{k}$, and final momentum, $\boldsymbol{k}+\boldsymbol{q}$, with respect to the $k_x$ axis (zigzag direction).  Also, we define the angle $\alpha$ for $\boldsymbol{q}$ and obtain
\begin{equation}
F_{ss^\prime}(\phi)=\frac{1}{2}\left[1+ss^\prime \frac{k \cos (2\theta) + q\cos(\theta+\alpha)}{\sqrt{k^2+q^2+2kq\cos(\theta-\alpha)}}\right].
\end{equation}
In this representation, $\theta$ is the angle over which $\boldsymbol{k}$ is integrated in Eq.~(\ref{eqn:cond}). 
The values of $s$ and $s^\prime$ are each either $+1$ or $-1$ which represents the upper (+) and lower (-) Dirac cones and for simplicity we will take the chemical potential to fall in the upper cone, representing an electron doped sheet. 
In general the conductivity, $\sigma_{ij}(q,\omega)$, can be written as a linear combination of a longitudinal part, $\sigma^L(q,\omega)$, and a transverse part, $\sigma^T(q,\omega)$.\cite{scholz:2011}  If we select $\boldsymbol{q}= q\hat{x}$, where $\alpha=0$ in our orientation, then one would find that the conductivity along the x-direction, $\sigma_{xx}(q,\omega)$, has only a contribution from the longitudinal conductivity (ie. $\sigma_{xx}(q,\omega)=\sigma^L(q,\omega)$) while the conductivity along the y-axis, $\sigma_{yy}(q,\omega)$, has only a contribution from the transverse part (ie.  $\sigma_{yy}(q,\omega)=\sigma^T(q,\omega)$).  Throughout this work, we are interested in examining quasiparticle-like peaks which appear only in the longitudinal part of the near-field optical spectra, and therefore will limit our discussion to examining $\sigma^L(q,\omega)$.

  Denoting the self energy by $\Sigma_s(k,\omega)$ the spectral function is
\begin{equation}
A^s(\boldsymbol{k},\omega)=\frac{1}{\pi} \frac{|{\rm Im}\Sigma_s(\boldsymbol{k},\omega)|}{[\omega- {\rm Re}\Sigma_s(\boldsymbol{k},\omega)- \epsilon_{{k}}^s]^2 +[{\rm Im}\Sigma_s(\boldsymbol{k},\omega)]^2},\label{eqn:akw}
\end{equation}
where $\epsilon_k^s=sv_Fk-\mu$ with $v_F$ the Fermi velocity. 
The self energy $\Sigma_s(\boldsymbol{k},\omega)$ can, in general, depend on band index and momentum as well as on energy.
Detailed calculations of the electron phonon interaction in graphene have been presented by Park and co-workers \cite{park:2007, park:2008} who conclude that a reasonable approximation to the complete calculations is to use a model of coupling to a single phonon of energy $\omega_E=200$~meV with no dependence on valley index and momentum. Here we follow this suggestion. If one were inclined, a distribution of phonon modes can be created through an integration over $\omega_E$ in Eq.~\ref{eqn:imsigma} weighted by the phonon density of states at each frequency, as has been done previously.\cite{dogan:2003,carbotte:2010}
 For an electron-phonon interaction (EPI) which includes coupling to a single phonon at frequency $\omega_E$, the imaginary part of the self energy, $\Sigma^{EPI}(\omega)$, is given by
\small
\begin{align}\label{eqn:imsigma}
&-{\rm Im}\Sigma^{EPI}(\omega ,\omega_E)=\nonumber\\
&\begin{cases}\displaystyle\frac{\pi {\rm A}}{W_c}|\omega-\omega_E+\mu_0|, & {\rm for}\quad \omega_E<\omega<W_c-\mu_0+\omega_E ,\\
           \displaystyle\frac{\pi {\rm A}}{W_c}|\omega+\omega_E+\mu_0|, & {\rm for}\quad -\omega_E>\omega>-W_c-\mu_0-\omega_E ,
\end{cases}
\end{align}
\normalsize
from which the real part can be obtained through a Kramers-Kronig transform.  We use the notation set out in Ref.~\onlinecite{carbotte:2010} where more details can be found.  In Eq.~(\ref{eqn:imsigma}), ${\rm A}$ is a constant that can be adjusted to get a desired value of the mass enhancement parameter, $\lambda$, and $W_c$ is a cut-off on the bare band energies adjusted to get the correct number of states in the Dirac approximation of two valleys in the Brillouin zone.  By definition $\lambda$ is obtained from the small $\omega$ limit of the real part of $\Sigma(\omega)$.  In this limit, ${\rm Re}\Sigma^{EPI}(\omega)={\rm Re}\Sigma^{EPI}(0)-\lambda \omega$ where the constant piece, ${\rm Re}\Sigma^{EPI}(0)$, shifts the chemical potential from its bare to interacting value.  

\begin{figure}
  \begin{center}
  \includegraphics[width=0.75\linewidth]{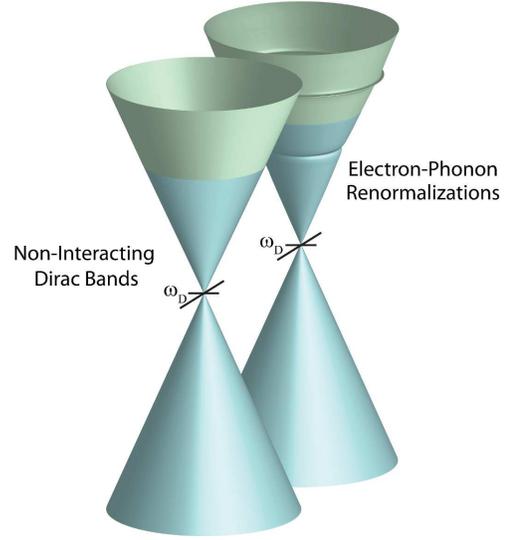}
  \end{center}
  \caption{\label{fig:2}(Color online) Schematic of doped Dirac bands.  An electron-phonon interaction renormalizes the Dirac bands, changing the slope through the Dirac point and producing kinks at $\omega=\pm \omega_E$ where here, the Fermi level ($\omega=0$) is illustrated by a change in color.  }
\end{figure}

The real part of the longitudinal conductivity follows by a numerical evaluation of Eq.(~\ref{eqn:cond}).  Results are shown in Fig.~\ref{fig:1} for ten values of $q/k_F$ from $0.0 \to 0.9$.  The solid black curve is the well known \cite{peres:2008, stauber:2008b,stauber:2008,carbotte:2010} $q=0$ case and is included here for comparison.  There is a Drude piece at small $\omega$ which is followed by a boson assisted part that sets in abruptly at $\omega=\omega_E$.  Here we have used a model for the self energy which also includes a constant impurity scattering term, $\eta$, in addition to the self energy from the electron-phonon interaction.  This constant scattering contributes no energy renormalization but broadens the Drude component. The EPI however, reduces the optical spectral weight of the coherent Drude part and redistributes it in the phonon assisted part which correspond to intraband absorption.  At higher energies (not shown here) there is additional absorption due to interband optical transitions which in the bare band case start abruptly at a frequency of twice the chemical potential.  With electron-phonon coupling these also spill into the Pauli blocked region below $2\mu$.  When $\boldsymbol{q}$ is finite we get the series of curves labelled by the energy of the position of the peaks in $\sigma^L(q,\omega)$.  One might refer to these peaks as quasiparticle peaks since their energies correspond exactly to $\epsilon_{k_F-q}^+$ in the bare band case.

Here for small values of $q$, the peaks in $\sigma(q,\omega)$ correspond instead to the renormalized quasiparticle energies $\epsilon^+(k_F-q)/[1+\lambda]$.  However, when $q$ approaches the phonon energy this simple renormalization ceases to hold. 
The broadening of the curves increases above $\omega=\omega_E$. This indicates that a new channel for scattering has opened; in our case, phonon emission.  In the inset of Fig.~\ref{fig:1}, we plot the imaginary part of the self energy, which contains a sharp jump in scattering at the phonon energy.
     Note that the Dirac point in the inset occurs where the imaginary part of the EPI self energy is zero, which comes from a zero in the electronic density of states at this energy. 

\begin{figure}
  \begin{center}
  \includegraphics[width=0.95\linewidth]{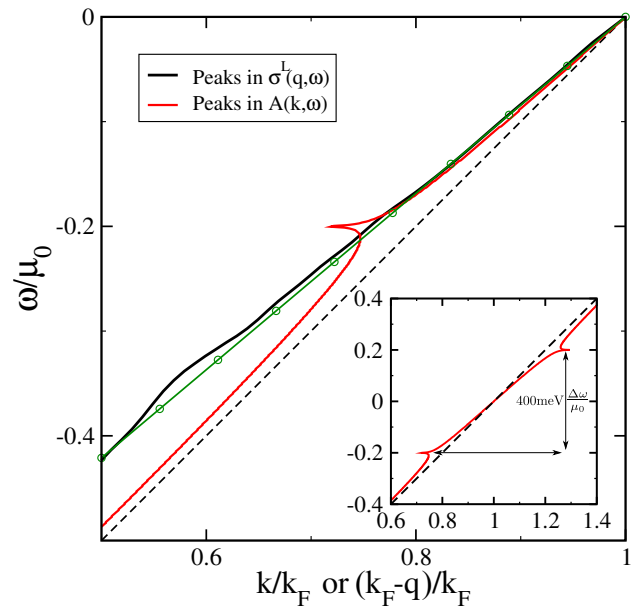}
  \end{center}
  \caption{\label{fig:3}(Color online) The peaks in $\sigma^L(q,\omega)$ plotted as a function of $k_F-q$ coincide with peaks in the EPI renormalized spectral function, $A(k,\omega)$, plotted as a function of $k$.  For comparison, the bare dispersion is shown by the dashed line, while the green line with circles is the bare dispersion divided by $(1+\lambda)$ which is the slope of the EPI renormalized spectral function (shown in red) at $k=k_F$.  The peaks in $\sigma^L(q,\omega)$ sample the renormalized dispersion until near $\omega=-\omega_E$.  Inset: $A(k,\omega)$ as in main frame but showing both positive and negative frequencies.  }
\end{figure}

 Fig.~\ref{fig:2} illustrates in a schematic fashion the effect of the electron-phonon interaction on the Dirac cone dispersion relations.  For the bare case, the Fermi velocity sets the angular dimension of each cone.  Interactions distort the cones in two important ways.  First of all, around the Fermi level the bare Fermi velocity is renormalized to a dressed value through a mass enhancement factor of $(1+\lambda)$.  As we move away from the Fermi energy this simple law starts to break down and phonon structures (referred to as `kinks') develop at $\omega=\pm\omega_E$ from the Fermi energy, and are measured in ARPES experiments.  These features are shown in the inset of Fig.~\ref{fig:3}. The solid red curve depicts the peaks in the phonon renormalized spectral function which illustrates the kinks at $\pm \omega_E$ as compared to the dashed black curve which is the bare cone dispersion.  In general the renormalized dispersion curves are also broadened.

\begin{figure}
  \begin{center}
  \includegraphics[width=0.8\linewidth]{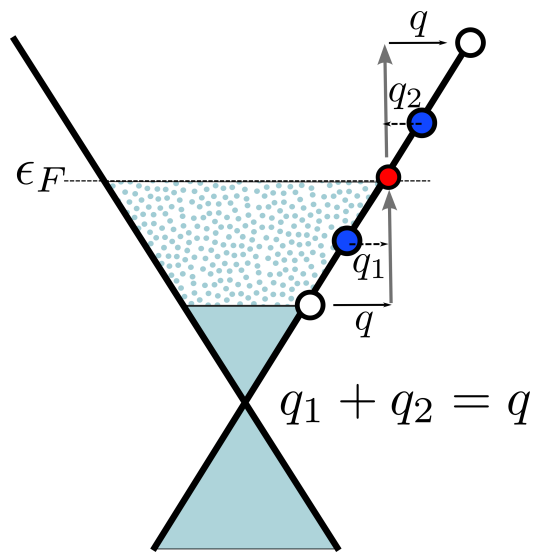}
  \end{center}
  \caption{\label{fig:new}(Color online) Schematic showing the many optical transitions from occupied states at energy $\epsilon=-q_1$ to unoccupied states at $\epsilon=q_2$ that all contribute to $\sigma^L(q,\omega)$ at $\omega=q$ when the dispersion curves are linear.  The large degeneracy of transitions gives a peak in the optical conductivity at $\omega=q$.  }
\end{figure}

\begin{figure}
  \begin{center}
  \includegraphics[width=0.9\linewidth]{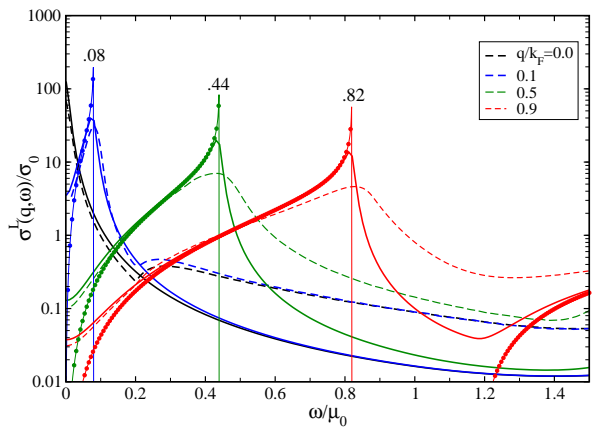}
  \end{center}
  \caption{\label{fig:4}(Color online) The real part of the finite momentum optical conductivity, $\sigma^L(q,\omega)/\sigma_0$ as a function of $\omega/\mu_0$ for a selection of values of $q/k_F$.  The dashed lines include the electron-phonon interaction identically to Fig.~\ref{fig:1} for $q/k_F$ displayed in the legend, while the solid lines include only the residual scattering of $\eta=0.005\mu_0$ but for shifted $q$ values (marked near the peaks).  The solid circles mark the analytical clean limit results, with an adjusted Fermi velocity to obtain the same peak positions.}
\end{figure} 

Here we want to know how the phonon `kinks' in the dressed dispersions manifest in the near field optics.  This is illustrated in the main frame of Fig.~\ref{fig:3} where the energies of the peaks in the real part of the optical conductivity $\sigma(q,\omega)$ are traced as a function of momentum, $k_F-q$, and are shown as the solid black curve.  These are to be compared with the solid red curve based on the peaks in the spectral density, $A(k,\omega)$ as a function of momentum, relative to the Fermi level.  The open circles with green curve represent a straight line having a Fermi velocity renormalized by $1/(1+\lambda)$ as compared to the bare dispersions shown as the dashed curve.  
Note that for energies a bit above the phonon energy at $\omega_E$ the peaks in the optical conductivity and in the single particle spectral density of ARPES   agree with each other and both fall on the renormalized Dirac fermion energy $\epsilon^+_{\boldsymbol{k}}/(1+\lambda)$ (light solid green curve).  In Fig.~\ref{fig:new} we show the transitions which contribute to the optical conductivity at $\omega=q$.  Because of the linearity of the Dirac dispersion curves, all transitions with $q_1+q_2 =q$ contributed to the same energy $\omega$ and this large degeneracy produces a peak at $\omega=q$ in $\sigma^L(q,\omega)$.  This only holds for linear dispersions.   Above $\omega_E$ this simple relationship ceases to hold with optics (solid black) roughly tracing the renormalized dispersion (green curve) while ARPES (solid red) is closer to the bare dispersion (dashed).
While the dressed quasiparticle spectral peaks show a phonon kink at $\omega=\omega_E$ there is no discernible structure in the optical conductivity at this energy.  
It is important to remember that optics involves a joint density of states from an initial occupied electron state to a final unoccupied one.  While many optical transitions correspond to a given energy it is clear that transitions from a kink to a second kink have a large weight and thus introduce structures into $\sigma^L(q,\omega)$ for frequencies which fall in magnitude just below $|2\omega_E|$. 

Very recently, Ashby and Carbotte \cite{ashby:2012} have studied the case of residual scattering alone and for small $q$ values have derived in first order an analytic formula which has proved quite accurate.  Here we have generalized their work to include the electron-phonon interaction in the regime where the renormalized quasiparticle energies are well represented by the formula $\epsilon^+_k/(1+\lambda)$.  We obtain
\begin{equation}\label{eqn:sigma1}
\frac{\sigma^L(q,\omega)}{\sigma_0}=\frac{4\mu_0}{\pi}\int\limits_0^{2\pi}\frac{d\theta}{2\pi}\frac{2\cos^2(\theta)(2\eta)}{(2\eta)^2+[\omega(1+\lambda)+q\cos\theta]^2}.
\end{equation}
To obtain this expression we have made an approximation by including only the coherent part of the Green's function, $G(k,\omega)$, where the standard $\omega$ is replaced by $\omega(1+\lambda)$.
In the clean limit ($\eta\to0$) we get
\begin{equation}\label{eqn:sigma2}
\frac{\sigma^L(q,\omega)}{\sigma_0}=\frac{8\mu_0}{\pi}\left( \frac{\omega}{\bar{q}}\right)^2 \frac{1}{\sqrt{\bar{q}^2-\omega^2}}\frac{1}{1+\lambda},
\end{equation}
with $\bar{q}=q/(1+\lambda)$.  We see that the square root singularity is at $\omega=\bar{q}$ rather than at $q$ as in the bare band case.  Also the optical spectral weight under $\sigma^L(q,\omega)$ is reduced by a factor of $1/(1+\lambda)$ and is given by
\begin{equation}\label{eqn:weight}
\int\limits_0^\infty d\omega \frac{\sigma^L(q,\omega)}{\sigma_0}= 2\mu_0/(1+\lambda).
\end{equation}
The missing optical spectral weight is transferred to a phonon assisted Holstein side band described by the incoherent piece of the Green's function. This band starts at the phonon energy, $\omega_E$, as is seen most clearly in the first two curves in Fig.~\ref{fig:1} for $q/k_F$= 0.0 and 0.08.
For a finite value of $\eta$, taking out the $(1+\lambda)$ factor next to $\omega$ in Eq.~(\ref{eqn:sigma1}) will leave an overall multiplicative factor of $1/(1+\lambda)$ in addition to changing $q$ to $\bar{q}$.  The scattering rate is also renormalized in the same manner as $\bar{\eta}=\eta/(1+\lambda)$.  This result means that the electron-phonon interaction reduces the effective residual scattering rate at low frequencies.  
These scalings are all in agreement with previous work\cite{carbotte:2010} for the conventional case of $q=0$ which exhibits a Drude peak (intraband transitions) in the energy region below the onset of the interband transitions which start at $\omega=2\mu_0$.

In Fig.~\ref{fig:3} we concentrated on the energies of the renormalized peaks in the optical conductivity and their relationship to  peaks in the quasiparticle spectral density.  Both are due to the real part of the quasiparticle self energies.  In Fig.~\ref{fig:4} we illustrate the effect of interactions on the broadening of the peaks which depends on the imaginary part of the self energy.  For each of the four values of $q/k_F$ we show 3 cases.  The dashed lines include both real and imaginary parts for the full self energy with coupling to the phonon as well as to impurities ($\eta$).  Open circles are for the clean limit (no scattering) of Eq.~(\ref{eqn:sigma2}) but where the energies $\epsilon_k^+$ of the bare bands have been shifted to peak at the same energy as the renormalized bands (values labelled on individual peaks).  
We first note that all of these curves have a rise as $\omega$ approaches $q$ from below, above which they drop sharply to zero.  The peak at $\omega=q$ in the solid lines, which include a small residual scattering rate of $\eta=0.005\mu_0$, shows smearing which results in a finite conductivity for $\omega$ just above $q$.  When the electron-phonon interaction is also included (dashed lines) this smearing of the spectral peak above $\omega=q$ is further increased due to boson assisted processes.  This additional interaction, however, has little effect at $\omega=q$ for small $q$ as illustrated in the $q/k_F=0$ and $0.1$ cases as compared with the larger $q=0.5$ and $0.9$.  For the small $q$ values there is no effect of the EPI in the region of the peak, except to shift the frequency at which the  peak occurs by a factor of the EPI renormalization, $(1+\lambda)$.  For larger $q$ values, the EPI causes a significant reduction in the peak height and filling of the Pauli blocked region.  These effects are due to a finite imaginary part of the electron-phonon self energy.  While the quasiparticle self energy has a sharp onset at $\omega=\omega_E$, as shown in the inset of Fig.~\ref{fig:1}, its effect on $\sigma^L(q,\omega)$ is much more gradual because many optical transitions (initial and final states) are involved in the creation of a hole-particle pair as depicted in Fig.~\ref{fig:new}.  If we restrict our discussion to scattering processes in the vicinity of $\omega=q$, then for $q<\omega_E$ these transitions will include initial and final states with energies always below $\omega_E$ and therefore not sample the EPI scattering rate.  For $q>\omega_E$ the transitions will begin to sample states with energies both above and below $|\omega_E|$ and therefore have an averaging of EPI scattering rates. 


To conclude, we have calculated the effect of electron-phonon coupling on the finite momentum optical response of graphene with the aim of getting a first understanding of how this interaction modifies the bare band results.  A Kubo formula is employed in the simplest bubble approximation which neglects vertex corrections, but includes electron-phonon renormalizations in the Dirac fermion spectral density $A(k,\omega)$.  In addition, there appears in the formula for the finite $\boldsymbol{q}$ conductivity an important scattering factor, $F_{ss^\prime}(\phi)$, which incorporates chiral properties of the Dirac charge carriers associated with initial and final electronic states.  As such, this factor depends in a fundamental way on the direction of $\boldsymbol{q}$ relative to the crystal lattice and direction of measurement.  The conductivity in the $x$-direction, $\sigma_{xx}$, can be decomposed for a general direction of momentum $\boldsymbol{q}$ into a linear combination of the longitudinal response for $q$ scattering which occurs along the zigzag direction ($k_x$) and a transverse part for $q$ scattering along the armchair direction ($k_y$).  Here, a peak is identified in $\sigma^L(q,\omega)$ which is directly related to the quasiparticle energy $\epsilon^+(k_F-q)$ in the bare band case.  The electron-phonon interaction is found to shift the position of this peak in energy and to broaden it.  Nevertheless, for values of $q$ small enough that only optical transitions involving both initial and final states with energies well below the phonon energy $\omega_E$ enter, the peaks in $\sigma^L(q,\omega)$ still track perfectly those in the electron spectral function $A(k,\omega)$ measured in ARPES and both relate directly to the renormalized quasiparticle energies $\epsilon^+(k_F-q)/(1+\lambda)$.  However, as the magnitude of $\boldsymbol{q}$ is increased, the quasiparticle dispersion curve starts to deviate significantly from linearity and exhibit a phonon kink.  At this point quasiparticle and optics begin to deviate from each other.  For example, optics shows no kink structure at $\omega=\omega_E$.  Instead the phonon structure in $\sigma^L(q,\omega)$ is found to have shifted to higher energies in the region of $\omega\lesssim2\omega_E$.  In this energy range, optical transitions involving a phonon kink in both initial and final states become possible and this leads to deviations of the optical spectrum from linearity in analogy to the phonon kinks at $\omega_E$ found in the renormalized Dirac fermion energies.  In this case there is a contribution from other optical transitions which emphasize less phonon structure and so the image of phonon is not as sharp in optics as it is in ARPES.  Finally, we found that at higher energies, $\omega>\omega_E$, the optical peaks continue to track well the renormalized energies $\epsilon^+(k_F-q)/(1+\lambda)$ while the quasiparticle peaks move instead towards their bare band values.


\begin{acknowledgments}
This research was supported in part by the Natural Sciences and
Engineering Research Council of Canada (NSERC) and the Canadian Institute
for Advanced Research (CIFAR). 
\end{acknowledgments}


\bibliographystyle{apsrev4-1}
\bibliography{bib}

\begin{thebibliography}{33}%
\makeatletter
\providecommand \@ifxundefined [1]{%
 \@ifx{#1\undefined}
}%
\providecommand \@ifnum [1]{%
 \ifnum #1\expandafter \@firstoftwo
 \else \expandafter \@secondoftwo
 \fi
}%
\providecommand \@ifx [1]{%
 \ifx #1\expandafter \@firstoftwo
 \else \expandafter \@secondoftwo
 \fi
}%
\providecommand \natexlab [1]{#1}%
\providecommand \enquote  [1]{``#1''}%
\providecommand \bibnamefont  [1]{#1}%
\providecommand \bibfnamefont [1]{#1}%
\providecommand \citenamefont [1]{#1}%
\providecommand \href@noop [0]{\@secondoftwo}%
\providecommand \href [0]{\begingroup \@sanitize@url \@href}%
\providecommand \@href[1]{\@@startlink{#1}\@@href}%
\providecommand \@@href[1]{\endgroup#1\@@endlink}%
\providecommand \@sanitize@url [0]{\catcode `\\12\catcode `\$12\catcode
  `\&12\catcode `\#12\catcode `\^12\catcode `\_12\catcode `\%12\relax}%
\providecommand \@@startlink[1]{}%
\providecommand \@@endlink[0]{}%
\providecommand \url  [0]{\begingroup\@sanitize@url \@url }%
\providecommand \@url [1]{\endgroup\@href {#1}{\urlprefix }}%
\providecommand \urlprefix  [0]{URL }%
\providecommand \Eprint [0]{\href }%
\providecommand \doibase [0]{http://dx.doi.org/}%
\providecommand \selectlanguage [0]{\@gobble}%
\providecommand \bibinfo  [0]{\@secondoftwo}%
\providecommand \bibfield  [0]{\@secondoftwo}%
\providecommand \translation [1]{[#1]}%
\providecommand \BibitemOpen [0]{}%
\providecommand \bibitemStop [0]{}%
\providecommand \bibitemNoStop [0]{.\EOS\space}%
\providecommand \EOS [0]{\spacefactor3000\relax}%
\providecommand \BibitemShut  [1]{\csname bibitem#1\endcsname}%
\let\auto@bib@innerbib\@empty
\bibitem [{\citenamefont {Orlita}\ and\ \citenamefont
  {Potemski}(2010)}]{orlita:2010}%
  \BibitemOpen
  \bibfield  {author} {\bibinfo {author} {\bibfnamefont {M.}~\bibnamefont
  {Orlita}}\ and\ \bibinfo {author} {\bibfnamefont {M.}~\bibnamefont
  {Potemski}},\ }\href@noop {} {\bibfield  {journal} {\bibinfo  {journal}
  {Semiconductor Science Technology}\ }\textbf {\bibinfo {volume} {25}},\
  \bibinfo {pages} {063001} (\bibinfo {year} {2010})}\BibitemShut {NoStop}%
\bibitem [{\citenamefont {Gusynin}\ \emph {et~al.}(2007)\citenamefont
  {Gusynin}, \citenamefont {Sharapov},\ and\ \citenamefont
  {Carbotte}}]{gusynin:2007}%
  \BibitemOpen
  \bibfield  {author} {\bibinfo {author} {\bibfnamefont {V.~P.}\ \bibnamefont
  {Gusynin}}, \bibinfo {author} {\bibfnamefont {S.~G.}\ \bibnamefont
  {Sharapov}}, \ and\ \bibinfo {author} {\bibfnamefont {J.~P.}\ \bibnamefont
  {Carbotte}},\ }\href@noop {} {\bibfield  {journal} {\bibinfo  {journal}
  {Phys. Rev. Lett.}\ }\textbf {\bibinfo {volume} {98}},\ \bibinfo {pages}
  {157402} (\bibinfo {year} {2007})}\BibitemShut {NoStop}%
\bibitem [{\citenamefont {Ando}\ \emph {et~al.}(2002)\citenamefont {Ando},
  \citenamefont {Zheng},\ and\ \citenamefont {Suzuura}}]{ando:2002}%
  \BibitemOpen
  \bibfield  {author} {\bibinfo {author} {\bibfnamefont {T.}~\bibnamefont
  {Ando}}, \bibinfo {author} {\bibfnamefont {Y.}~\bibnamefont {Zheng}}, \ and\
  \bibinfo {author} {\bibfnamefont {H.}~\bibnamefont {Suzuura}},\ }\href@noop
  {} {\bibfield  {journal} {\bibinfo  {journal} {J. Phys. Soc. Jpn.}\ }\textbf
  {\bibinfo {volume} {71}},\ \bibinfo {pages} {1318} (\bibinfo {year}
  {2002})}\BibitemShut {NoStop}%
\bibitem [{\citenamefont {Gusynin}\ \emph {et~al.}(2009)\citenamefont
  {Gusynin}, \citenamefont {Sharapov},\ and\ \citenamefont
  {Carbotte}}]{gusynin:2009}%
  \BibitemOpen
  \bibfield  {author} {\bibinfo {author} {\bibfnamefont {V.~P.}\ \bibnamefont
  {Gusynin}}, \bibinfo {author} {\bibfnamefont {S.~G.}\ \bibnamefont
  {Sharapov}}, \ and\ \bibinfo {author} {\bibfnamefont {J.~P.}\ \bibnamefont
  {Carbotte}},\ }\href@noop {} {\bibfield  {journal} {\bibinfo  {journal} {New
  Journal of Physics}\ }\textbf {\bibinfo {volume} {11}},\ \bibinfo {pages}
  {095013} (\bibinfo {year} {2009})}\BibitemShut {NoStop}%
\bibitem [{\citenamefont {Li}\ \emph {et~al.}(2008)\citenamefont {Li},
  \citenamefont {Henriksen}, \citenamefont {Jiang}, \citenamefont {Hao},
  \citenamefont {Martin}, \citenamefont {Kim}, \citenamefont {Stormer},\ and\
  \citenamefont {Basov}}]{li:2008}%
  \BibitemOpen
  \bibfield  {author} {\bibinfo {author} {\bibfnamefont {Z.}~\bibnamefont
  {Li}}, \bibinfo {author} {\bibfnamefont {E.~A.}\ \bibnamefont {Henriksen}},
  \bibinfo {author} {\bibfnamefont {Z.}~\bibnamefont {Jiang}}, \bibinfo
  {author} {\bibfnamefont {Z.}~\bibnamefont {Hao}}, \bibinfo {author}
  {\bibfnamefont {M.~C.}\ \bibnamefont {Martin}}, \bibinfo {author}
  {\bibfnamefont {P.}~\bibnamefont {Kim}}, \bibinfo {author} {\bibfnamefont
  {H.~L.}\ \bibnamefont {Stormer}}, \ and\ \bibinfo {author} {\bibfnamefont
  {D.~N.}\ \bibnamefont {Basov}},\ }\href@noop {} {\bibfield  {journal}
  {\bibinfo  {journal} {Nat. Phys.}\ }\textbf {\bibinfo {volume} {4}},\
  \bibinfo {pages} {532} (\bibinfo {year} {2008})}\BibitemShut {NoStop}%
\bibitem [{\citenamefont {Peres}\ \emph {et~al.}(2008)\citenamefont {Peres},
  \citenamefont {Stauber},\ and\ \citenamefont {{Castro Neto}}}]{peres:2008}%
  \BibitemOpen
  \bibfield  {author} {\bibinfo {author} {\bibfnamefont {N.~M.~R.}\
  \bibnamefont {Peres}}, \bibinfo {author} {\bibfnamefont {T.}~\bibnamefont
  {Stauber}}, \ and\ \bibinfo {author} {\bibfnamefont {A.~H.}\ \bibnamefont
  {{Castro Neto}}},\ }\href@noop {} {\bibfield  {journal} {\bibinfo  {journal}
  {EPL}\ }\textbf {\bibinfo {volume} {84}},\ \bibinfo {pages} {38002} (\bibinfo
  {year} {2008})}\BibitemShut {NoStop}%
\bibitem [{\citenamefont {Stauber}\ and\ \citenamefont
  {Peres}(2008)}]{stauber:2008b}%
  \BibitemOpen
  \bibfield  {author} {\bibinfo {author} {\bibfnamefont {T.}~\bibnamefont
  {Stauber}}\ and\ \bibinfo {author} {\bibfnamefont {N.~M.~R.}\ \bibnamefont
  {Peres}},\ }\href@noop {} {\bibfield  {journal} {\bibinfo  {journal} {J.
  Phys.: Condens. Matter}\ }\textbf {\bibinfo {volume} {20}},\ \bibinfo {pages}
  {055002} (\bibinfo {year} {2008})}\BibitemShut {NoStop}%
\bibitem [{\citenamefont {Stauber}\ \emph {et~al.}(2008)\citenamefont
  {Stauber}, \citenamefont {Peres},\ and\ \citenamefont {{Castro
  Neto}}}]{stauber:2008}%
  \BibitemOpen
  \bibfield  {author} {\bibinfo {author} {\bibfnamefont {T.}~\bibnamefont
  {Stauber}}, \bibinfo {author} {\bibfnamefont {N.~M.~R.}\ \bibnamefont
  {Peres}}, \ and\ \bibinfo {author} {\bibfnamefont {A.~H.}\ \bibnamefont
  {{Castro Neto}}},\ }\href@noop {} {\bibfield  {journal} {\bibinfo  {journal}
  {Phys. Rev. B}\ }\textbf {\bibinfo {volume} {78}},\ \bibinfo {pages} {085418}
  (\bibinfo {year} {2008})}\BibitemShut {NoStop}%
\bibitem [{\citenamefont {Carbotte}\ \emph {et~al.}(2010)\citenamefont
  {Carbotte}, \citenamefont {Nicol},\ and\ \citenamefont
  {Sharapov}}]{carbotte:2010}%
  \BibitemOpen
  \bibfield  {author} {\bibinfo {author} {\bibfnamefont {J.~P.}\ \bibnamefont
  {Carbotte}}, \bibinfo {author} {\bibfnamefont {E.~J.}\ \bibnamefont {Nicol}},
  \ and\ \bibinfo {author} {\bibfnamefont {S.~G.}\ \bibnamefont {Sharapov}},\
  }\href@noop {} {\bibfield  {journal} {\bibinfo  {journal} {Phys. Rev. B}\
  }\textbf {\bibinfo {volume} {81}},\ \bibinfo {pages} {045419} (\bibinfo
  {year} {2010})}\BibitemShut {NoStop}%
\bibitem [{\citenamefont {Prange}\ and\ \citenamefont
  {Kadanoff}(1964)}]{prange:1964}%
  \BibitemOpen
  \bibfield  {author} {\bibinfo {author} {\bibfnamefont {R.~E.}\ \bibnamefont
  {Prange}}\ and\ \bibinfo {author} {\bibfnamefont {L.~P.}\ \bibnamefont
  {Kadanoff}},\ }\href@noop {} {\bibfield  {journal} {\bibinfo  {journal}
  {Phys. Rev.}\ }\textbf {\bibinfo {volume} {134}},\ \bibinfo {pages} {A566}
  (\bibinfo {year} {1964})}\BibitemShut {NoStop}%
\bibitem [{\citenamefont {Mori}\ \emph {et~al.}(2008)\citenamefont {Mori},
  \citenamefont {Nicol}, \citenamefont {Shiizuka}, \citenamefont {Kuniyasu},
  \citenamefont {Nojima}, \citenamefont {Toyota},\ and\ \citenamefont
  {Carbotte}}]{mori:2008}%
  \BibitemOpen
  \bibfield  {author} {\bibinfo {author} {\bibfnamefont {T.}~\bibnamefont
  {Mori}}, \bibinfo {author} {\bibfnamefont {E.~J.}\ \bibnamefont {Nicol}},
  \bibinfo {author} {\bibfnamefont {S.}~\bibnamefont {Shiizuka}}, \bibinfo
  {author} {\bibfnamefont {K.}~\bibnamefont {Kuniyasu}}, \bibinfo {author}
  {\bibfnamefont {T.}~\bibnamefont {Nojima}}, \bibinfo {author} {\bibfnamefont
  {N.}~\bibnamefont {Toyota}}, \ and\ \bibinfo {author} {\bibfnamefont {J.~P.}\
  \bibnamefont {Carbotte}},\ }\href@noop {} {\bibfield  {journal} {\bibinfo
  {journal} {Phys. Rev. B}\ }\textbf {\bibinfo {volume} {77}},\ \bibinfo
  {pages} {174515} (\bibinfo {year} {2008})}\BibitemShut {NoStop}%
\bibitem [{\citenamefont {Li}\ \emph {et~al.}(2009)\citenamefont {Li},
  \citenamefont {Luican},\ and\ \citenamefont {Andrei}}]{li:2009}%
  \BibitemOpen
  \bibfield  {author} {\bibinfo {author} {\bibfnamefont {G.}~\bibnamefont
  {Li}}, \bibinfo {author} {\bibfnamefont {A.}~\bibnamefont {Luican}}, \ and\
  \bibinfo {author} {\bibfnamefont {E.~Y.}\ \bibnamefont {Andrei}},\
  }\href@noop {} {\bibfield  {journal} {\bibinfo  {journal} {Phys. Rev. Lett.}\
  }\textbf {\bibinfo {volume} {102}},\ \bibinfo {pages} {176804} (\bibinfo
  {year} {2009})}\BibitemShut {NoStop}%
\bibitem [{\citenamefont {Mitrovic}\ and\ \citenamefont
  {Carbotte}(1983{\natexlab{a}})}]{mitrovic:1983}%
  \BibitemOpen
  \bibfield  {author} {\bibinfo {author} {\bibfnamefont {B.}~\bibnamefont
  {Mitrovic}}\ and\ \bibinfo {author} {\bibfnamefont {J.~P.}\ \bibnamefont
  {Carbotte}},\ }\href@noop {} {\bibfield  {journal} {\bibinfo  {journal} {Can.
  J. Phys.}\ }\textbf {\bibinfo {volume} {61}},\ \bibinfo {pages} {758}
  (\bibinfo {year} {1983}{\natexlab{a}})}\BibitemShut {NoStop}%
\bibitem [{\citenamefont {Mitrovic}\ and\ \citenamefont
  {Carbotte}(1983{\natexlab{b}})}]{mitrovic:1983:2}%
  \BibitemOpen
  \bibfield  {author} {\bibinfo {author} {\bibfnamefont {B.}~\bibnamefont
  {Mitrovic}}\ and\ \bibinfo {author} {\bibfnamefont {J.~P.}\ \bibnamefont
  {Carbotte}},\ }\href@noop {} {\bibfield  {journal} {\bibinfo  {journal} {Can.
  J. Phys.}\ }\textbf {\bibinfo {volume} {61}},\ \bibinfo {pages} {784}
  (\bibinfo {year} {1983}{\natexlab{b}})}\BibitemShut {NoStop}%
\bibitem [{\citenamefont {Cappelluti}\ and\ \citenamefont
  {Pietronero}(2003)}]{cappelluti:2003}%
  \BibitemOpen
  \bibfield  {author} {\bibinfo {author} {\bibfnamefont {E.}~\bibnamefont
  {Cappelluti}}\ and\ \bibinfo {author} {\bibfnamefont {L.}~\bibnamefont
  {Pietronero}},\ }\href@noop {} {\bibfield  {journal} {\bibinfo  {journal}
  {Phys. Rev. B}\ }\textbf {\bibinfo {volume} {68}},\ \bibinfo {pages} {224511}
  (\bibinfo {year} {2003})}\BibitemShut {NoStop}%
\bibitem [{\citenamefont {Nicol}\ and\ \citenamefont
  {Carbotte}(2009)}]{nicol:2009}%
  \BibitemOpen
  \bibfield  {author} {\bibinfo {author} {\bibfnamefont {E.~J.}\ \bibnamefont
  {Nicol}}\ and\ \bibinfo {author} {\bibfnamefont {J.~P.}\ \bibnamefont
  {Carbotte}},\ }\href@noop {} {\bibfield  {journal} {\bibinfo  {journal}
  {Phys. Rev. B}\ }\textbf {\bibinfo {volume} {80}},\ \bibinfo {pages}
  {081415(R)} (\bibinfo {year} {2009})}\BibitemShut {NoStop}%
\bibitem [{\citenamefont {Bostwick}\ \emph {et~al.}(2007)\citenamefont
  {Bostwick}, \citenamefont {Ohta}, \citenamefont {Seyller}, \citenamefont
  {Horn},\ and\ \citenamefont {Rotenberg}}]{bostwick:2007}%
  \BibitemOpen
  \bibfield  {author} {\bibinfo {author} {\bibfnamefont {A.}~\bibnamefont
  {Bostwick}}, \bibinfo {author} {\bibfnamefont {T.}~\bibnamefont {Ohta}},
  \bibinfo {author} {\bibfnamefont {T.}~\bibnamefont {Seyller}}, \bibinfo
  {author} {\bibfnamefont {K.}~\bibnamefont {Horn}}, \ and\ \bibinfo {author}
  {\bibfnamefont {E.}~\bibnamefont {Rotenberg}},\ }\href@noop {} {\bibfield
  {journal} {\bibinfo  {journal} {Nat. Phys.}\ }\textbf {\bibinfo {volume}
  {3}},\ \bibinfo {pages} {36} (\bibinfo {year} {2007})}\BibitemShut {NoStop}%
\bibitem [{\citenamefont {Zhou}\ \emph {et~al.}(2008)\citenamefont {Zhou},
  \citenamefont {Siegel}, \citenamefont {Fedorov},\ and\ \citenamefont
  {Lanzara}}]{zhou:2008}%
  \BibitemOpen
  \bibfield  {author} {\bibinfo {author} {\bibfnamefont {S.~Y.}\ \bibnamefont
  {Zhou}}, \bibinfo {author} {\bibfnamefont {D.~A.}\ \bibnamefont {Siegel}},
  \bibinfo {author} {\bibfnamefont {A.~V.}\ \bibnamefont {Fedorov}}, \ and\
  \bibinfo {author} {\bibfnamefont {A.}~\bibnamefont {Lanzara}},\ }\href@noop
  {} {\bibfield  {journal} {\bibinfo  {journal} {Phys. Rev. B}\ }\textbf
  {\bibinfo {volume} {78}},\ \bibinfo {pages} {193404} (\bibinfo {year}
  {2008})}\BibitemShut {NoStop}%
\bibitem [{\citenamefont {Calandra}\ and\ \citenamefont
  {Mauri}(2007)}]{calandra:2007}%
  \BibitemOpen
  \bibfield  {author} {\bibinfo {author} {\bibfnamefont {M.}~\bibnamefont
  {Calandra}}\ and\ \bibinfo {author} {\bibfnamefont {F.}~\bibnamefont
  {Mauri}},\ }\href@noop {} {\bibfield  {journal} {\bibinfo  {journal} {Phys.
  Rev. B}\ }\textbf {\bibinfo {volume} {76}},\ \bibinfo {pages} {205411}
  (\bibinfo {year} {2007})}\BibitemShut {NoStop}%
\bibitem [{\citenamefont {Park}\ \emph
  {et~al.}(2009{\natexlab{a}})\citenamefont {Park}, \citenamefont {Giustino},
  \citenamefont {Spataru}, \citenamefont {Cohen},\ and\ \citenamefont
  {Louie}}]{park:2009}%
  \BibitemOpen
  \bibfield  {author} {\bibinfo {author} {\bibfnamefont {C.-H.}\ \bibnamefont
  {Park}}, \bibinfo {author} {\bibfnamefont {F.}~\bibnamefont {Giustino}},
  \bibinfo {author} {\bibfnamefont {C.~D.}\ \bibnamefont {Spataru}}, \bibinfo
  {author} {\bibfnamefont {M.~L.}\ \bibnamefont {Cohen}}, \ and\ \bibinfo
  {author} {\bibfnamefont {S.~G.}\ \bibnamefont {Louie}},\ }\href@noop {}
  {\bibfield  {journal} {\bibinfo  {journal} {Phys. Rev. Lett.}\ }\textbf
  {\bibinfo {volume} {102}},\ \bibinfo {pages} {076803} (\bibinfo {year}
  {2009}{\natexlab{a}})}\BibitemShut {NoStop}%
\bibitem [{\citenamefont {Park}\ \emph
  {et~al.}(2009{\natexlab{b}})\citenamefont {Park}, \citenamefont {Giustino},
  \citenamefont {Spataru}, \citenamefont {Cohen},\ and\ \citenamefont
  {Louie}}]{park:2009:nl}%
  \BibitemOpen
  \bibfield  {author} {\bibinfo {author} {\bibfnamefont {C.-H.}\ \bibnamefont
  {Park}}, \bibinfo {author} {\bibfnamefont {F.}~\bibnamefont {Giustino}},
  \bibinfo {author} {\bibfnamefont {C.~D.}\ \bibnamefont {Spataru}}, \bibinfo
  {author} {\bibfnamefont {M.~L.}\ \bibnamefont {Cohen}}, \ and\ \bibinfo
  {author} {\bibfnamefont {S.~G.}\ \bibnamefont {Louie}},\ }\href@noop {}
  {\bibfield  {journal} {\bibinfo  {journal} {Nano Lett.}\ }\textbf {\bibinfo
  {volume} {9}},\ \bibinfo {pages} {4234} (\bibinfo {year}
  {2009}{\natexlab{b}})}\BibitemShut {NoStop}%
\bibitem [{\citenamefont {Tse}\ and\ \citenamefont {{Das
  Sarma}}(2007)}]{tse:2007}%
  \BibitemOpen
  \bibfield  {author} {\bibinfo {author} {\bibfnamefont {W.-K.}\ \bibnamefont
  {Tse}}\ and\ \bibinfo {author} {\bibfnamefont {S.}~\bibnamefont {{Das
  Sarma}}},\ }\href@noop {} {\bibfield  {journal} {\bibinfo  {journal} {Phys.
  Rev. Lett.}\ }\textbf {\bibinfo {volume} {99}},\ \bibinfo {pages} {236802}
  (\bibinfo {year} {2007})}\BibitemShut {NoStop}%
\bibitem [{\citenamefont {Fei}\ \emph {et~al.}(2011)\citenamefont {Fei},
  \citenamefont {Andreev}, \citenamefont {Bao}, \citenamefont {Zhang},
  \citenamefont {McLeod}, \citenamefont {Wang}, \citenamefont {Stewart},
  \citenamefont {Zhao}, \citenamefont {Dominguez}, \citenamefont {Thiemens},
  \citenamefont {Fogler}, \citenamefont {Tauber}, \citenamefont {Castro-Neto},
  \citenamefont {Lau}, \citenamefont {Keilmann},\ and\ \citenamefont
  {Basov}}]{fei:2011}%
  \BibitemOpen
  \bibfield  {author} {\bibinfo {author} {\bibfnamefont {Z.}~\bibnamefont
  {Fei}}, \bibinfo {author} {\bibfnamefont {G.~O.}\ \bibnamefont {Andreev}},
  \bibinfo {author} {\bibfnamefont {W.}~\bibnamefont {Bao}}, \bibinfo {author}
  {\bibfnamefont {L.~M.}\ \bibnamefont {Zhang}}, \bibinfo {author}
  {\bibfnamefont {A.~S.}\ \bibnamefont {McLeod}}, \bibinfo {author}
  {\bibfnamefont {C.}~\bibnamefont {Wang}}, \bibinfo {author} {\bibfnamefont
  {M.~K.}\ \bibnamefont {Stewart}}, \bibinfo {author} {\bibfnamefont
  {Z.}~\bibnamefont {Zhao}}, \bibinfo {author} {\bibfnamefont {G.}~\bibnamefont
  {Dominguez}}, \bibinfo {author} {\bibfnamefont {M.}~\bibnamefont {Thiemens}},
  \bibinfo {author} {\bibfnamefont {M.~M.}\ \bibnamefont {Fogler}}, \bibinfo
  {author} {\bibfnamefont {M.~J.}\ \bibnamefont {Tauber}}, \bibinfo {author}
  {\bibfnamefont {A.~H.}\ \bibnamefont {Castro-Neto}}, \bibinfo {author}
  {\bibfnamefont {C.~N.}\ \bibnamefont {Lau}}, \bibinfo {author} {\bibfnamefont
  {F.}~\bibnamefont {Keilmann}}, \ and\ \bibinfo {author} {\bibfnamefont
  {D.~N.}\ \bibnamefont {Basov}},\ }\href@noop {} {\bibfield  {journal}
  {\bibinfo  {journal} {Nano Letters}\ }\textbf {\bibinfo {volume} {11}},\
  \bibinfo {pages} {4701} (\bibinfo {year} {2011})}\BibitemShut {NoStop}%
\bibitem [{\citenamefont {Fei}\ \emph {et~al.}(2012)\citenamefont {Fei},
  \citenamefont {Rodin}, \citenamefont {Andreev}, \citenamefont {Bao},
  \citenamefont {McLeod}, \citenamefont {Wagner}, \citenamefont {Zhang},
  \citenamefont {Zhao}, \citenamefont {Dominguez}, \citenamefont {Thiemens},
  \citenamefont {Fogler}, \citenamefont {Castro-Neto}, \citenamefont {Lau},
  \citenamefont {Keilmann},\ and\ \citenamefont {Basov}}]{fei:2012}%
  \BibitemOpen
  \bibfield  {author} {\bibinfo {author} {\bibfnamefont {Z.}~\bibnamefont
  {Fei}}, \bibinfo {author} {\bibfnamefont {S.}~\bibnamefont {Rodin}}, \bibinfo
  {author} {\bibfnamefont {G.~O.}\ \bibnamefont {Andreev}}, \bibinfo {author}
  {\bibfnamefont {W.}~\bibnamefont {Bao}}, \bibinfo {author} {\bibfnamefont
  {A.~S.}\ \bibnamefont {McLeod}}, \bibinfo {author} {\bibfnamefont
  {M.}~\bibnamefont {Wagner}}, \bibinfo {author} {\bibfnamefont {L.~M.}\
  \bibnamefont {Zhang}}, \bibinfo {author} {\bibfnamefont {Z.}~\bibnamefont
  {Zhao}}, \bibinfo {author} {\bibfnamefont {G.}~\bibnamefont {Dominguez}},
  \bibinfo {author} {\bibfnamefont {M.}~\bibnamefont {Thiemens}}, \bibinfo
  {author} {\bibfnamefont {M.~M.}\ \bibnamefont {Fogler}}, \bibinfo {author}
  {\bibfnamefont {A.~H.}\ \bibnamefont {Castro-Neto}}, \bibinfo {author}
  {\bibfnamefont {C.~N.}\ \bibnamefont {Lau}}, \bibinfo {author} {\bibfnamefont
  {F.}~\bibnamefont {Keilmann}}, \ and\ \bibinfo {author} {\bibfnamefont
  {D.~N.}\ \bibnamefont {Basov}},\ }\href@noop {} {\bibfield  {journal}
  {\bibinfo  {journal} {Nature}\ }\textbf {\bibinfo {volume} {487}},\ \bibinfo
  {pages} {82} (\bibinfo {year} {2012})}\BibitemShut {NoStop}%
\bibitem [{\citenamefont {Chen}\ \emph {et~al.}(2012)\citenamefont {Chen},
  \citenamefont {Badioli}, \citenamefont {Alonso-Gonz\'alez}, \citenamefont
  {Thongrattanasiri}, \citenamefont {Huth}, \citenamefont {Osmond},
  \citenamefont {Spasenovi\'c}, \citenamefont {Centeno}, \citenamefont
  {Pesquera}, \citenamefont {Godignon}, \citenamefont {Elorza}, \citenamefont
  {Camara}, \citenamefont {{de Abajo}}, \citenamefont {hillenbrand},\ and\
  \citenamefont {Koppens}}]{chen:2012}%
  \BibitemOpen
  \bibfield  {author} {\bibinfo {author} {\bibfnamefont {J.}~\bibnamefont
  {Chen}}, \bibinfo {author} {\bibfnamefont {M.}~\bibnamefont {Badioli}},
  \bibinfo {author} {\bibfnamefont {P.}~\bibnamefont {Alonso-Gonz\'alez}},
  \bibinfo {author} {\bibfnamefont {S.}~\bibnamefont {Thongrattanasiri}},
  \bibinfo {author} {\bibfnamefont {F.}~\bibnamefont {Huth}}, \bibinfo {author}
  {\bibfnamefont {J.}~\bibnamefont {Osmond}}, \bibinfo {author} {\bibfnamefont
  {M.}~\bibnamefont {Spasenovi\'c}}, \bibinfo {author} {\bibfnamefont
  {A.}~\bibnamefont {Centeno}}, \bibinfo {author} {\bibfnamefont
  {A.}~\bibnamefont {Pesquera}}, \bibinfo {author} {\bibfnamefont
  {P.}~\bibnamefont {Godignon}}, \bibinfo {author} {\bibfnamefont {A.~Z.}\
  \bibnamefont {Elorza}}, \bibinfo {author} {\bibfnamefont {N.}~\bibnamefont
  {Camara}}, \bibinfo {author} {\bibfnamefont {F.~J.~G.}\ \bibnamefont {{de
  Abajo}}}, \bibinfo {author} {\bibfnamefont {R.}~\bibnamefont {hillenbrand}},
  \ and\ \bibinfo {author} {\bibfnamefont {F.~H.~L.}\ \bibnamefont {Koppens}},\
  }\href@noop {} {\bibfield  {journal} {\bibinfo  {journal} {Nature}\ }\textbf
  {\bibinfo {volume} {487}},\ \bibinfo {pages} {77} (\bibinfo {year}
  {2012})}\BibitemShut {NoStop}%
\bibitem [{\citenamefont {Carbotte}\ \emph {et~al.}(2012)\citenamefont
  {Carbotte}, \citenamefont {LeBlanc},\ and\ \citenamefont
  {Nicol}}]{carbotte:2012}%
  \BibitemOpen
  \bibfield  {author} {\bibinfo {author} {\bibfnamefont {J.~P.}\ \bibnamefont
  {Carbotte}}, \bibinfo {author} {\bibfnamefont {J.~P.~F.}\ \bibnamefont
  {LeBlanc}}, \ and\ \bibinfo {author} {\bibfnamefont {E.~J.}\ \bibnamefont
  {Nicol}},\ }\href@noop {} {\bibfield  {journal} {\bibinfo  {journal} {Phys.
  Rev. B}\ }\textbf {\bibinfo {volume} {85}},\ \bibinfo {pages} {201411(R)}
  (\bibinfo {year} {2012})}\BibitemShut {NoStop}%
\bibitem [{\citenamefont {Lundqvist}(1967)}]{lundqvist:1967}%
  \BibitemOpen
  \bibfield  {author} {\bibinfo {author} {\bibfnamefont {B.}~\bibnamefont
  {Lundqvist}},\ }\href@noop {} {\bibfield  {journal} {\bibinfo  {journal}
  {Phys. Kondens. Materie}\ }\textbf {\bibinfo {volume} {6}},\ \bibinfo {pages}
  {193} (\bibinfo {year} {1967})}\BibitemShut {NoStop}%
\bibitem [{\citenamefont {Cappelluti}\ and\ \citenamefont
  {Benfatto}(2009)}]{cappelluti:2009}%
  \BibitemOpen
  \bibfield  {author} {\bibinfo {author} {\bibfnamefont {E.}~\bibnamefont
  {Cappelluti}}\ and\ \bibinfo {author} {\bibfnamefont {L.}~\bibnamefont
  {Benfatto}},\ }\href@noop {} {\bibfield  {journal} {\bibinfo  {journal}
  {Phys. Rev. B}\ }\textbf {\bibinfo {volume} {79}},\ \bibinfo {pages} {035419}
  (\bibinfo {year} {2009})}\BibitemShut {NoStop}%
\bibitem [{\citenamefont {Scholz}\ and\ \citenamefont
  {Schliemann}(2011)}]{scholz:2011}%
  \BibitemOpen
  \bibfield  {author} {\bibinfo {author} {\bibfnamefont {A.}~\bibnamefont
  {Scholz}}\ and\ \bibinfo {author} {\bibfnamefont {J.}~\bibnamefont
  {Schliemann}},\ }\href@noop {} {\bibfield  {journal} {\bibinfo  {journal}
  {Phys. Rev. B}\ }\textbf {\bibinfo {volume} {83}},\ \bibinfo {pages} {235409}
  (\bibinfo {year} {2011})}\BibitemShut {NoStop}%
\bibitem [{\citenamefont {Park}\ \emph {et~al.}(2007)\citenamefont {Park},
  \citenamefont {Giustino}, \citenamefont {Cohen},\ and\ \citenamefont
  {Louie}}]{park:2007}%
  \BibitemOpen
  \bibfield  {author} {\bibinfo {author} {\bibfnamefont {C.-H.}\ \bibnamefont
  {Park}}, \bibinfo {author} {\bibfnamefont {F.}~\bibnamefont {Giustino}},
  \bibinfo {author} {\bibfnamefont {M.~L.}\ \bibnamefont {Cohen}}, \ and\
  \bibinfo {author} {\bibfnamefont {S.~G.}\ \bibnamefont {Louie}},\ }\href@noop
  {} {\bibfield  {journal} {\bibinfo  {journal} {Phys. Rev. Lett.}\ }\textbf
  {\bibinfo {volume} {99}},\ \bibinfo {pages} {086804} (\bibinfo {year}
  {2007})}\BibitemShut {NoStop}%
\bibitem [{\citenamefont {Park}\ \emph {et~al.}(2008)\citenamefont {Park},
  \citenamefont {Giustino}, \citenamefont {Spataru}, \citenamefont {Cohen},\
  and\ \citenamefont {Louie}}]{park:2008}%
  \BibitemOpen
  \bibfield  {author} {\bibinfo {author} {\bibfnamefont {C.-H.}\ \bibnamefont
  {Park}}, \bibinfo {author} {\bibfnamefont {F.}~\bibnamefont {Giustino}},
  \bibinfo {author} {\bibfnamefont {C.~D.}\ \bibnamefont {Spataru}}, \bibinfo
  {author} {\bibfnamefont {M.~L.}\ \bibnamefont {Cohen}}, \ and\ \bibinfo
  {author} {\bibfnamefont {S.~G.}\ \bibnamefont {Louie}},\ }\href@noop {}
  {\bibfield  {journal} {\bibinfo  {journal} {Nano Lett.}\ }\textbf {\bibinfo
  {volume} {8}},\ \bibinfo {pages} {4229} (\bibinfo {year} {2008})}\BibitemShut
  {NoStop}%
\bibitem [{\citenamefont {Do\v{g}an}\ and\ \citenamefont
  {Marsiglio}(2003)}]{dogan:2003}%
  \BibitemOpen
  \bibfield  {author} {\bibinfo {author} {\bibfnamefont {F.}~\bibnamefont
  {Do\v{g}an}}\ and\ \bibinfo {author} {\bibfnamefont {F.}~\bibnamefont
  {Marsiglio}},\ }\href@noop {} {\bibfield  {journal} {\bibinfo  {journal}
  {Phys. Rev. B}\ }\textbf {\bibinfo {volume} {68}},\ \bibinfo {pages} {165102}
  (\bibinfo {year} {2003})}\BibitemShut {NoStop}%
\bibitem [{\citenamefont {Ashby}\ and\ \citenamefont
  {Carbotte}(2012)}]{ashby:2012}%
  \BibitemOpen
  \bibfield  {author} {\bibinfo {author} {\bibfnamefont {P.~E.~C.}\
  \bibnamefont {Ashby}}\ and\ \bibinfo {author} {\bibfnamefont {J.~P.}\
  \bibnamefont {Carbotte}},\ }\href@noop {} {\bibfield  {journal} {\bibinfo
  {journal} {Phys. Rev. B}\ }\textbf {\bibinfo {volume} {86}},\ \bibinfo
  {pages} {165405} (\bibinfo {year} {2012})}\BibitemShut {NoStop}%
\end{thebibliography}%

\end{document}